 \documentclass[12pt,english]{article}
 \usepackage{babel}
 \usepackage{amsmath,mathrsfs,amssymb}
 \usepackage{graphicx,placeins,caption2}
 \usepackage{indentfirst}
 
 \setcaptionwidth{\textwidth}
 
 \newcommand{\beq}[1]{\begin{equation}\label{#1}}
 \newcommand{\eeq}{\end{equation}}
 \newcommand{\bea}[1]{\begin{eqnarray}\label{#1}}
 \newcommand{\eea}{\end{eqnarray}}

\begin{document}
 
\title{Geodesic Motions in AdS Soliton Background Space-time}
\author{Han-qing Shi and Ding-fang Zeng\footnote{correspondence author}\\
\it Theoretical Physics Division, College of Applied Sciences\\
\it Beijing University of Technology, Beijing, China
\\
\footnotesize 631834083@qq.com and dfzeng@bjut.edu.cn}

\date{\today}
\maketitle

\abstract{
We study both massive and massless particle's geodesic motion in the background of general dimensional AdS-Sol space-time. We find that the massive particles oscillate along the radial direction, while massless particles experience one-time bouncing as they approach the ``horizon'' line of the soliton. Our results provide a direct way to understand the negative energy/masses leading to the AdS-Sol geometry. As a potential application, we extend the point particle to a 3-brane and fix the background as a 5+1 dimension AdS-Sol, thus obtain a very natural bouncing/cyclic cosmological model.
}

\tableofcontents

\section{Introduction}

Comparing with the Anti deSitter Black Hole (AdS-BH), the history of AdS Soliton  (AdS-Sol) is very short. It's first appearance is in G. T. Horowitz and R. C. Myers (HM) exploration \cite{adsSoliton} of AdS/CFT (abbr. for Anti deSitter/Conformal Field theory) correspondences in which the CFT or gauge theory side has supersymmetry-breaking boundary conditions. HM find that this geometry has negative energy, which happens to correspond to the negative Casimir energy in the gauge theory side. The stability of the gauge theory motivates HM to make a new positive energy conjecture, according to which AdS-Sol is the lowest negative energy solution among all solutions with the same asymptotic behavious. After the work of HM, Galloway, Surya, Woolgar prove \cite{uniqueTheorem} a uniqueness theorem for the AdS-Sol geometry thus offer a significant support for HM's new positive energy conjecture.

The negative energy feature of AdS-Sol  geometry motivate many studies on phase transitions related with the AdS-BHs. For example, even before the proof of \cite{uniqueTheorem}, S. Surya, K. Schleich and D. M. Witt \cite{phTransConfDConf} showed that, although no phase transition exists between the AdS-BH with Ricci-flat horizon and thermal-AdS geometry, it would occur between such black holes and the AdS-Sol geometry. In the AdS/CFT duality, cold, large black holes correspond to deconfining phases of the gauge theory system; while hot, small AdS-Sol's correspond to confining phases. After \cite{phTransConfDConf}, this transition is widely explored in many contexts, firstly aiming at realistic AdS/QCD (quantum chromodynamics) models building \cite{RGCai07024, eChargedPhTrans, RGCai070533, RGCai0707, conDeconTrans, phTransAdSZk}, laterly at the AdS/CMP (condensed matter physics), especially insulator/superconductor model explorations \cite{phTransAdSsoliton, marginStableModes, magEffphTrans, hairySolitonAdS5, higherDimensions, spin2condensates, holoHeatTransport, adsSoLconfscalarhair, adsSoLinstability, HNUcollab}, and some, such as \cite{enEntropySOLBH}, focuses on the entanglement entropy questions related with the phase transitions involving AdS-Sol geometries. In all these explorations, the negative energy feature of the AdS-Sol geometry plays key roles. So a more intuitive understanding of this feature is necessary. This paper will provide such a try by studying the geodesic motion of both massive and massless particles in this background space time.

The organisation of this paper is as follows. This section is a short introduction to the motivation initiating this work. The next section will review the standard definition and calculations of the energy/masses generating the AdS-Sol geometry. Section \ref{secMassive} and \ref{secMassless} study geodesic motions of massive and massless particles respectively. Section \ref{secCosmo} provide a potential cosmological application of the results of section \ref{secMassive} and \ref{secMassless}. The last section is the conclusion of the paper.

\section{Energy/masses generating the AdS-Sol geometry}
It is a very instructive way in studying the physics of a space-time geometry by considering geodesic motion of test particles in it. References \cite{adsBHgeodesic} and \cite{ctcAdSbh} are two typical works in this aspects. We will start from the math form of $n+1$ dimensional AdS-BH and AdS-Sol geometries in poincare coordinates,
\beq{}
AdS-BH:~ds^2=\frac{r^2}{\ell^2}(-hd\eta^2+d\vec{x}\cdot d\vec{x}+dy^2)+\frac{\ell^2dr^2}{r^2h}
\label{bhMetric}
\eeq
\beq{}
AdS-Sol:~ds^2=\frac{r^2}{\ell^2}(+hdy^2+d\vec{x}\cdot d\vec{x}-d\eta^2)+\frac{\ell^2dr^2}{r^2h}
\label{solMetric}
\eeq
\beq{}
h=1-\frac{r_0^n}{r^n}
\eeq
from which we can easily see that the latter is the double-wick rotation of the former;
while to avoid conical singularities in the $i\eta-r$ plane in AdS-BH case and $y-r$ plane in the AdS-Sol case, the time coordinate $\eta$ in AdS-BH and the space coordinate $y$ in AdS-Sol should be identified with periods
\beq{}
i\eta\sim i\eta+\beta,~y\sim y+\beta,~\beta=\frac{4\pi}{h'_h}=\frac{4\pi r_0}{n}
\label{periodsCoordinate}
\eeq
respectively.
\begin{figure}[h]
\begin{center}
\includegraphics[scale=1]{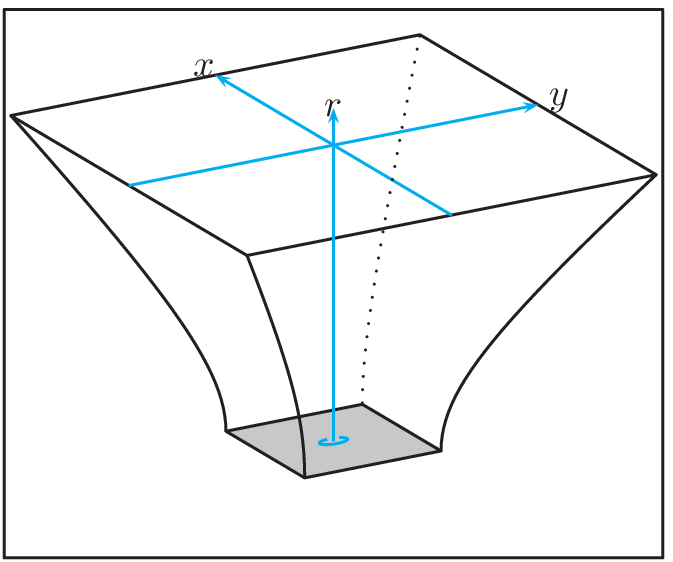}
\rule{3mm}{0pt}
\includegraphics[scale=1]{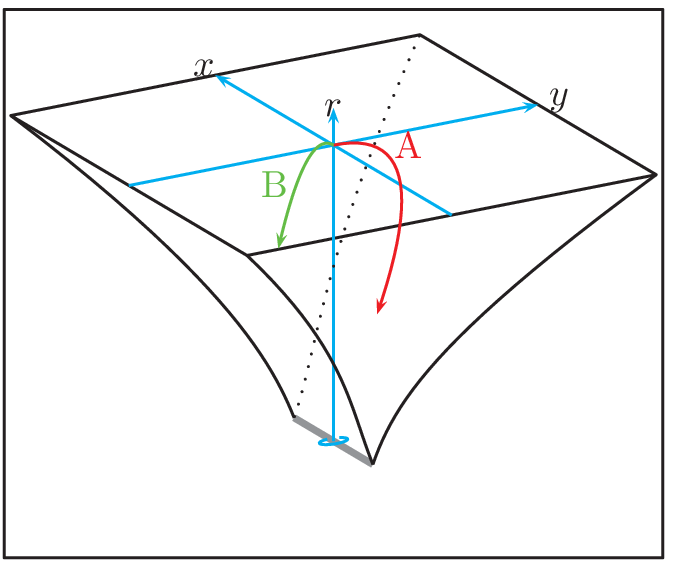}
\par
\end{center}
\caption{Intuitive pictures of AdS-BH(left) and AdS-Sol(right) geometry. In the AdS-Sol background, we also plot two kinds of time-like or light-like geodesic lines. The first is in the $y-r$ plane, while the second is in $x-r$ plane.}
\label{figGeometry}
\end{figure}
From the metric expression, we easily see that the horizon surface of AdS-BHs has topologies of the form $R^{n-1}$. As comparisons, the $r=r_0$ ``line'' is not the horizon of AdS-Sol space-time at all, due to its inability to hide the singular surface $r=0$ of the geometry. For the solitons, $r=r_0$ ``line'' has the topology of the form $R^{n-2,1}$. This difference has been portrayed in figure \ref{figGeometry} schematically.

In general relativities, the definition and calculation of energy/masses leading to some space-time geometry are always relative to some referring geometry with time-translation invariance (For AdS-Sol and AdS-BH's, this is just the simple AdS space-time),
\beq{}
E=-\frac{1}{8\pi G_N}\int N(K-K_0)
\label{enDefinition}
\eeq
where $K$ is the trace of extrinsic curvature of a space-like hyper-surface inside which the object we are interested in lives, $K_0$ is that of the hyper-surface with the same intrinsic geometry in the referring space-time, $N$ is the modulus of a time-like Killing fields on the hyper-surface, the integration should be done along that hyper-surface. In this paper's space-time, we will take the equal-$r$($r\rightarrow\infty$) surface as the enwrapping surface. Obviously, they are parametrised by the $\{\vec{x},y\}$ coordinate. In this case, the Killing vector is a constant proportional to $\frac{r}{\ell}$ everywhere on the surface, while the integration ${\displaystyle\int} K$ over the surface simplifies to the normal derivative of the area of the hyper-surface at all
\beq{}
\int K=n^\mu\partial_\mu A,~\int K_0=n_0^\mu\partial_\mu A_0
\label{NKtonpA}
\eeq
Corresponding to the metric form \eqref{bhMetric} and $\eqref{solMetric}$, $n^\mu=\delta^\mu_{~r}/\sqrt{g_{rr}}$ is the same to both of them, but
\beq{}
A_{bh}=\frac{r^{n-1}}{\ell^{n-1}}L_x^{n-2}L_y,~(n^\mu\partial_\mu A)_{bh}=\frac{(n-1)r^{n-1}}{\ell^n}\big(1-\frac{r_0^n}{r^n}\big)^\frac{1}{2}
L_x^{n-2}L_y
\label{bhNKIntegration}
\eeq
\beq{}
A_{sol}=\frac{r^{n-1}}{\ell^{n-1}}\sqrt{h}L_x^{n-2}L_y,~(n^\mu\partial_\mu A)_{sol}=\Big[\frac{(n-1)r^{n-1}}{\ell^n}\big(1-\frac{r_0^n}{r^n}\big)^\frac{1}{2}+\frac{r^n}{\ell^n}\big(\frac{+nr_0^n}{2r^{n+1}}\big)\Big]L_x^{n-2}L_y
\label{solNKIntegration}
\eeq
where $L_x$ and $L_y$ represent the extension of the surface on $x$ and $y$ directions. In the AdS-bh, both $L_x$ and $L_y$ is infinite in principle. While in the AdS-Sol case, $L_y$ should be identified periodically as equation \eqref{periodsCoordinate} implies. However, when we focus on energy/mass densities, neither $L_x$ nor $L_y$'s concrete value matters. According to calculations above \eqref{enDefinition}-\eqref{solNKIntegration}, we know
\beq{}
\frac{E}{L_x^{n-2}L_y}\Big|_{bh}=-\frac{r}{\ell}\frac{(n-1)r^{n-1}}{\ell^n}\Big[\sqrt{1-\frac{r_0^n}{r^n}}-\sqrt{1-\frac{0}{r^n}}\Big]_{r\rightarrow\infty}=\frac{(n-1)r_0^n}{2\ell^{n+1}}
\eeq
\beq{}
\frac{E}{L_x^{n-2}L_y}\Big|_{sol}=\frac{E}{L_x^{n-2}L_y}\Big|_{bh}
+\Big(-\frac{r}{\ell}\Big)\frac{r^n}{\ell^n}\frac{nr_0^n}{2r^{n+1}}
=-\frac{r_0^n}{2\ell^{n+1}}
\eeq
Obviously, this calculation tells us that the AdS-sol space-time is generated by some negative energy sources, but it tells us nothing more about this strange energy form, see also reference \cite{energyTension}. The only physics about this form of energy is, it may be related with the negative zero point energy of the dual gauge field theory. However, just as we will see below, even in the pure gravity side, this negative energy has direct observatory effects, it repel ordinary particles both with and without masses. That is, the AdS-Sol anti-gravitates. Basing on results of this paper, we will be able to explain the energy of AdS-Sol's in the same sense as in usual field theory monopoles  \cite{ftSolitonReview}.
\FloatBarrier

\section{Geodesic motion of massive particles}
\label{secMassive}
From this section on, we will focus on AdS-Sol geometry exclusively instead of parallelly with AdS-BH's. In this section, we will pay attention on two kinds of time-like geodesic motions in the AdS-Sol background space time. They are parametrized as follows
\beq{}
A)~\{\eta=\eta(\lambda),~\vec{x}\equiv0,~y=y(\lambda),~r=r(\lambda)\}
\rule{4mm}{0pt}
\label{aGeoAnsatz}
\eeq
\beq{}
B)~\{\eta=\eta(\lambda),~x^i=x(\lambda)\delta^i_1,~y\equiv0,~r=r(\lambda)\}
\label{bGeoAnsatz}
\eeq
They are displayed in figure \ref{figGeometry} and labeled with A and B respectively. The function form of all $x^\mu(\lambda)$'s is determined by minimalsation of the world-line action $S=\int\sqrt{-\dot{x}^\mu\dot{x}^\nu g_{\mu\nu}}d\lambda$. We will use world-line reparametrisation invariance to set $\eta(\lambda)=\lambda$ in both cases\footnote{We will rename $\lambda$ to $\eta$ when necessary}, so that
\beq{}
S_A=\int\sqrt{1-h\dot{y}^2-\frac{\ell^4\dot{r}^2}{r^4h}}\frac{r(\lambda)}{\ell}d\lambda
\equiv\int\mathcal{L}_A[y,r]d\lambda
\label{actionA}
\eeq
\beq{}
S_B=\int\sqrt{1-\dot{x}^2-\frac{\ell^4\dot{r}^2}{r^4h}}\frac{r(\lambda)}{\ell}d\lambda
\equiv\int\mathcal{L}_B[x,r]d\lambda
\label{actionB}
\eeq
The corresponding equations of motion have the form
\bea{}
A)~~\frac{-h\dot{y}}{\sqrt{1-h\dot{y}^2-\frac{\ell\dot{r}^2}{r^4h}}}\frac{r}{\ell}=C_1,~
\frac{1}{\sqrt{1-h\dot{y}^2-\cdots}}\frac{r}{\ell}=C_2
\label{eqYr0}
\eea
\bea{}
B)~~\frac{-\dot{x}}{\sqrt{1-\dot{x}^2-\frac{\ell\dot{r}^2}{r^4h}}}\frac{r}{\ell}=C'_1,~
\frac{1}{\sqrt{1-\dot{x}^2-\cdots}}\frac{r}{\ell}=C'_2
\rule{7mm}{0pt}
\label{eqXr0}
\eea
with $C_1(C_1')$, $C_2(C'_2)$ being integration constants which are positively correlated with the initial velocity in $y$(or $x$)- direction and the total energies of the particles respectively. 

\begin{figure*}[ht]
\begin{center}
\parbox{0.43\textwidth}{
\includegraphics[scale=0.6]{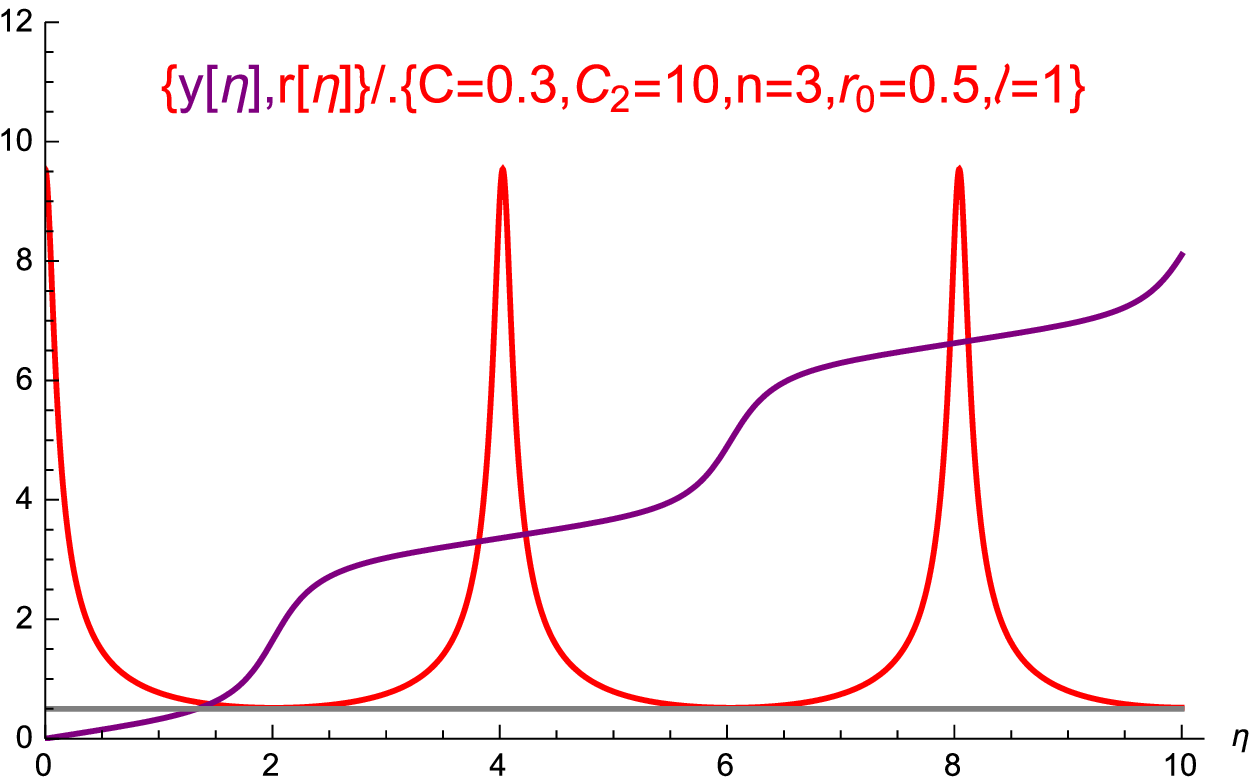}
\\
\includegraphics[scale=0.6]{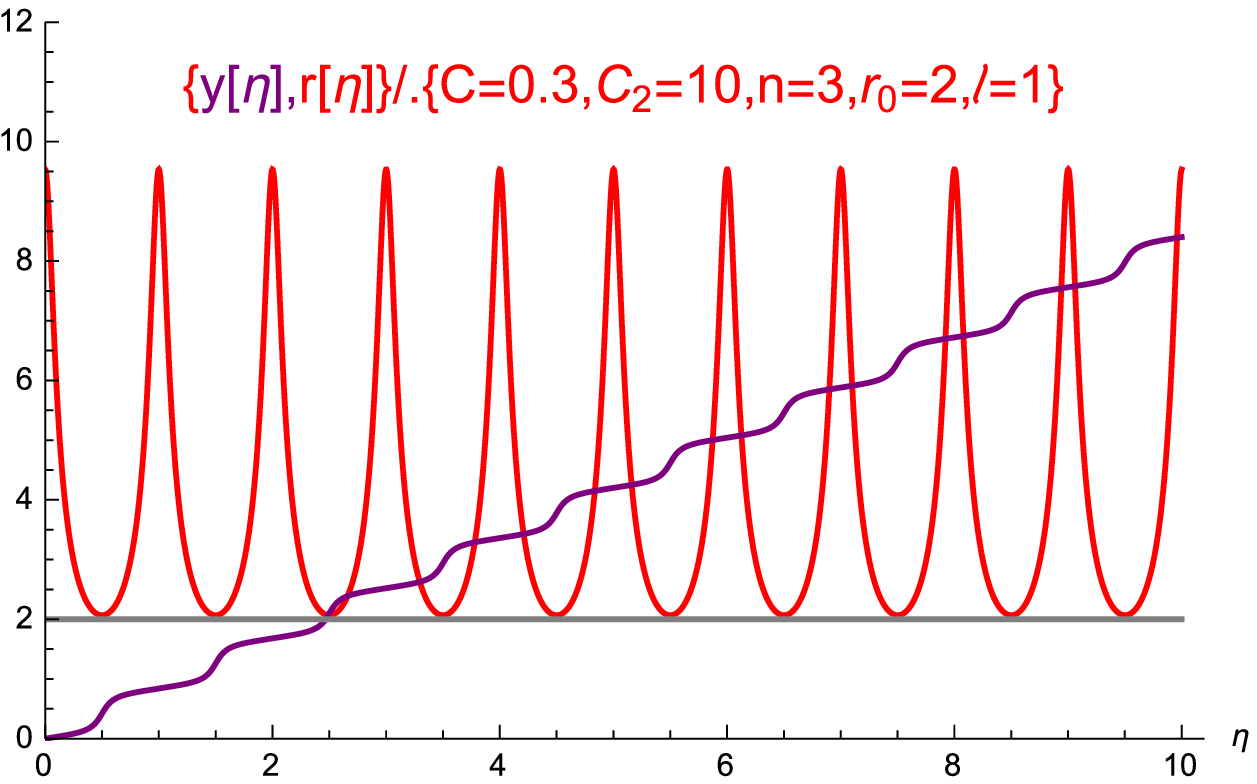}
\\
\rule{2mm}{0pt}\includegraphics[scale=1]{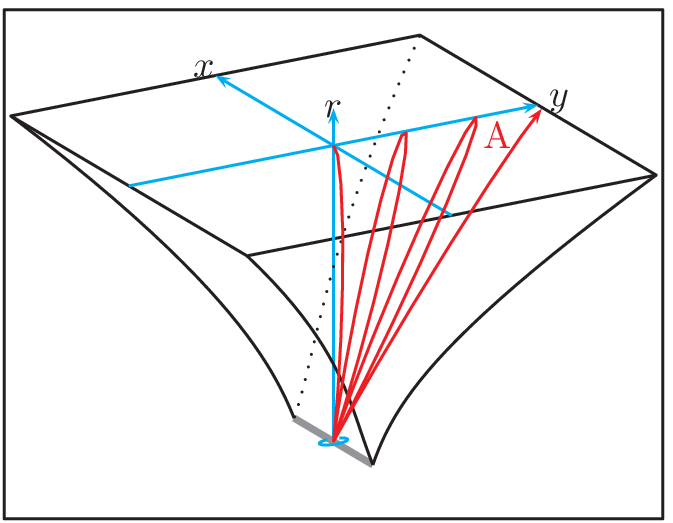}
}\parbox{0.43\textwidth}{
\includegraphics[scale=0.6]{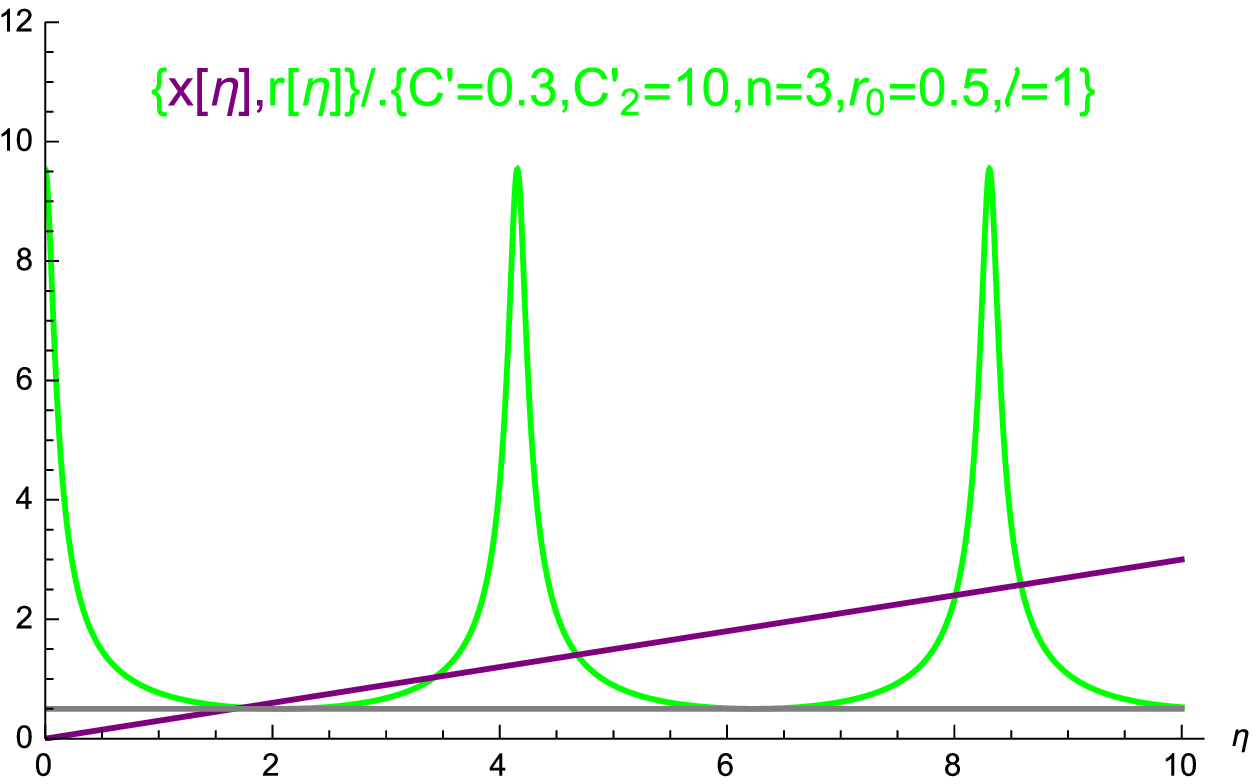}
\\
\includegraphics[scale=0.6]{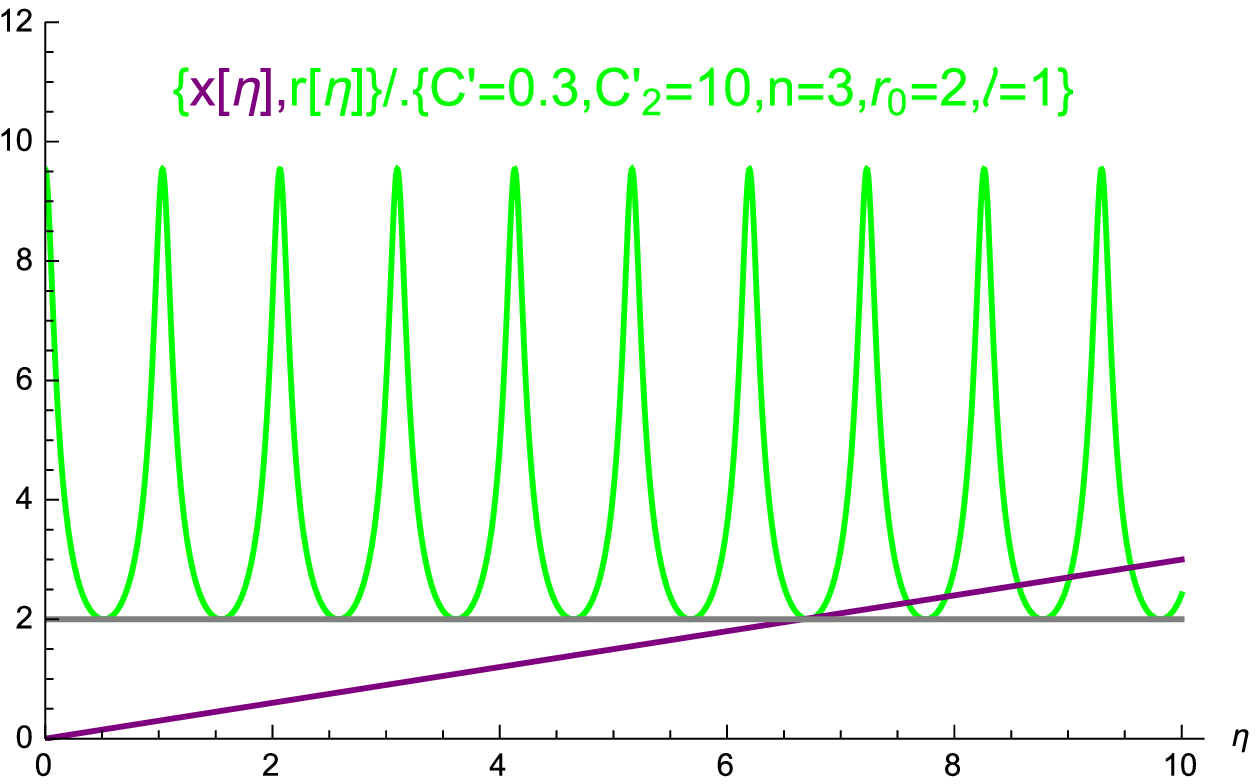}
\\
\rule{2mm}{0pt}\includegraphics[scale=1]{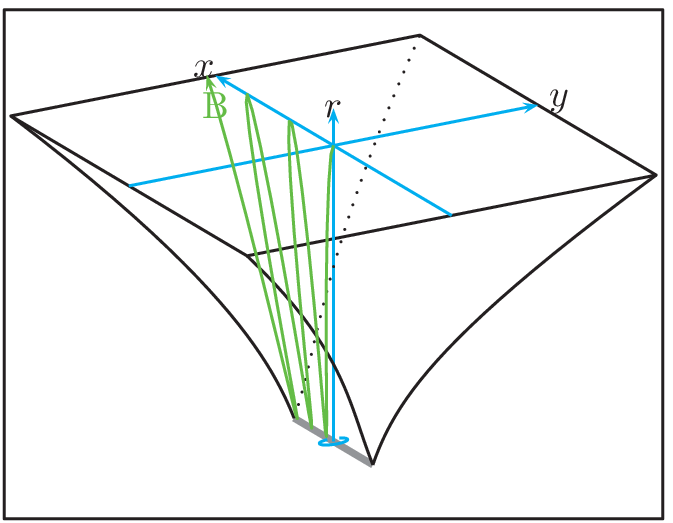}
}
\caption{left, numeric solutions to the equation of motion \eqref{eqYr1} of massive particles in AdS-Sol background \eqref{solMetric} with initial conditions \eqref{icy}. The corresponding particles move in the $y-r$ plane of the space-time. The monotonically increasing function is $y(\eta)$ while the oscillatory one is $r(\eta)$.  Right, numeric solutions to \eqref{eqXr1}+\eqref{icx}, the relevant particles moving in the $x-r$ plane. The simple linear function is $x(\eta)$ while $r(\eta)$ oscillates. In all figures, the horizontal gray line is the position of ``horizon''-line. All parameters involved has been displayed in the figure.}
\label{figMfunction}
\end{center}
\end{figure*}

Equations \eqref{eqYr0} and \eqref{eqXr0} could be written into forms more convenient for numeric calculations
\bea{}
h[r]\dot{y}=-C,~2\ddot{r}=&&\hspace{-5mm}4r^3(1-\frac{C^2}{h})-(4-n)r^{3-n}r_0^n(1-\frac{C^2}{h})-\frac{6r^5}{C_2^2\ell^2}+\frac{(6-n)r^{5-n}r_0^n}{C_2^2\ell^2}
\label{eqYr1}
\\
&&\hspace{-7mm}+r^4\big(+\frac{C^2}{h^2}h'\big)-r^{4-n}r_0^n\big(+\frac{C^2}{h^2}h'\big)
,~C\equiv\frac{C_1}{C_2}<1
\nonumber
\eea
\bea{}
\dot{x}=-C',~2\ddot{r}=&&\hspace{-5mm}4r^3(1-{C'}^2)-(4-n)r^{3-n}r_0^n(1-{C'}^2)-\frac{6r^5}{{C'_2}^2\ell^2}+\frac{(6-n)r^{5-n}r_0^n}{{C'}_2^2\ell^2}
\label{eqXr1}
\\
&&\hspace{-5mm}C'\equiv\frac{C_1'}{C_2'}<1
\nonumber
\eea
where $h'$ is the derivative of $h(r)$ with respect to $r$. The latter part of this two equations comes from differentiations of $\dot{r}^2$ expression solved from equations \eqref{eqYr0} and \eqref{eqXr0}. Although this differentiating operation makes order of the goal equations higher. It makes numerical calculations more fluent because in the higher order equation, we need not manually decide when/where to change the sign of $\dot{r}=\pm\sqrt{F[r,\cdots]}$ as $r$ varies. However, in the higher order equations, the initial condition must be set properly so that the first order equations be hold consistently. For example, the following conditions 
\beq{}
i.c.A):~ y[0]=0,~r[0]=C_2\ell\sqrt{1-C^2},~\dot{r}[0]=0
\label{icy}
\eeq
\beq{}
i.c.B):~ x[0]=0,~r[0]=C_2\ell\sqrt{1-C^2},~\dot{r}[0]=0
\label{icx}
\eeq
are consistent. While conditions of the follow form (with $\dot{r}_0$ being some arbitrary initial speed along the $r$-direction)
\beq{}
i.c'.A):~ y[0]=0,~r[0]=C_2\ell\sqrt{1-C^2},~\dot{r}[0]=\dot{r}_0
\label{icyerr}
\eeq
\beq{}
i.c'.B):~ x[0]=0,~r[0]=C_2\ell\sqrt{1-C^2},~\dot{r}[0]=\dot{r}_0
\label{icxerr}
\eeq
will be inconsistent. If one neglect the fact that equations \eqref{eqYr1} and \eqref{eqXr1} comes from the differentiation of first order equations \eqref{eqYr0} and \eqref{eqXr0}, one may think that the initial conditions \eqref{icyerr} and \eqref{icxerr} are totally reasonable. In fact using them as initial conditions to solve equation \eqref{eqYr1} and \eqref{eqXr1} will yield unphysical results.

Figure \ref{figMfunction} displays numeric solutions to the equation of motion \eqref{eqYr1} under the initial condition \eqref{icy} and the equation of motion \eqref{eqXr1} with the initial condition \eqref{icx}. The most attracting feature of these numerical solution is, the test particle's vertical motion $r(\eta)$ has oscillatory behaviour, while the horizontal motion $x(\eta)$ and $y(\eta)$ are always towards fixed directions as time elapses. That is, as a test particle moves towards the ``horizon''-line of the soliton, it receives a repelling forces, while as it moves too far away from the ``horizon''-line, it receives attracting forces. The former originates from the negative energy/masses thus anti-gravity of the soliton, the latter comes from the potential-well effect of asymptotic AdS property of the space-time. This phenomena indicates that, without resorting to dual gauge theory interpretation, the negative energy of AdS-sol geometry could give us direct observatory effects in pure gravity framework.

Qualitatively, figure \ref{figMfunction} also tells us that the oscillation period is positively related with the mass/size $r_0$ of the AdS-Sol. More massive or bigger soliton tends to drive the test particle to make faster oscillatory motion. Numeric excercises tell us, in all $n+1$($n\geqslant3$) dimensional AdS-Sol space-time, $r_0$ is the most relevant one affecting the periods of test particle's oscillation among the four relevant parameters $\{C\equiv\frac{C_1}{C_2}$, $C_2$, $n$, $r_0\}$. $\ell$ is irrelevant because we can always set it to $1$ through appropriate length-unit choice. In principle, this fact should be interpretable from the equation of motion \eqref{eqYr1} or \eqref{eqXr1}, which are essentially nonlinear oscillatory equations. However, in practices, this is somewhat complicated due to our inability to analytically solve general higher than 5-degree algebraic equations to get the oscillator's equivalent position and effective stiffness coefficient (in $n+1=4+1$ dimension case, things may become solvable). For this reason and for its deviation from our main interests, we will not dwell on this point anymore.
\FloatBarrier

\section{Massless particles}
\label{secMassless}
For massless particles, we will still focus on the two kinds of geodesic motion of the form \eqref{aGeoAnsatz} and \eqref{bGeoAnsatz}. The point is, in this time, the relevant motion function are no longer determined by the actions \eqref{actionA} and \eqref{actionB}. Instead, they are determined by 
the following two sets of equations of motion
\beq{}
A)~~\dot{\eta}^2-h\dot{y}^2-\frac{\ell^4\dot{r}^2}{r^4h}=0,~~\dot{y}r^2h=C_y,~~\frac{C_y^2/\ell^4+\dot{r}^2}{h}=C_r^2
\label{eqYrph1}
\eeq
\beq{}
B)~~\dot{\eta}^2-\dot{x}^2-\frac{\ell^4\dot{r}^2}{r^4h}=0,~~\dot{x}r^2=C_x,~~\frac{\dot{r}^2}{h}={C'_r}^{_{\scriptstyle2}}
\rule{6mm}{0pt}
\label{eqXrph1}
\eeq
where $C_y$, $C_x$, $C_r$ and $C'_r$ are four integration constants, over-dots represent derivatives with respect to $\lambda$, while $h'\equiv\frac{dh(r)}{dr}$.
The first part of both this two equations come from the light-like feature of the mass-less particle's orbit. While the latter part of them is the integration of relevant components of standard geodesic equation 
\beq{}
\ddot{x}^\mu+\Gamma^\mu_{\mu\nu}\dot{x}^\mu\dot{x^\nu}=0
\eeq
with the relevant nonzero Christopher connection components given by
\beq{}
\Gamma^y_{~yr}=\Gamma^y_{~ry}=\frac{1}{r}+\frac{h'}{2h}
,~\Gamma^x_{~xr}=\Gamma^x_{~rx}=\frac{1}{r}
\eeq
\beq{}
\Gamma^r_{~\eta\eta}=\frac{r^3}{\ell^4}h,~\Gamma^r_{~xx}=-\frac{r^3}{\ell^4}h
\eeq
\beq{}
\Gamma^r_{~yy}=-\frac{r^3}{\ell^4}\big(h^2+\frac{r}{2}hh'\big),~
\Gamma^r_{~rr}=-\big(\frac{1}{r}+\frac{h'}{2h}\big)
\eeq

Analytically solving equations \eqref{eqYrph1} and \eqref{eqXrph1} is almost impossible. So like the case of massive particles, we still resort to numeric methods. To avoid manual sign-decision of $\dot{r}=\pm\sqrt{\cdots}$ in numerics, we first derive from them the expression of $\dot{r}^2$, then differentiating the results with with respect to $\lambda$ so that they become second order equations again 
\beq{}
A)~~\ddot{r}=\frac{{C_r'}^{_{\scriptstyle2}}}{2}h'
,~~
B)~~\ddot{r}=\frac{C_r^2}{2}h'
\eeq
This two equation looks the same. However, they come from different first order differential equations. For this reason, we need use different initial conditions for them in numerics
\beq{}
i.c.A)~~y[0]=0,~r[0]=r_\mathrm{ini},~\dot{r}[0]=-\Big[C_r^2h-C_y^2/\ell^4\Big]^\frac{1}{2}_{r=_\mathrm{ini}}
\label{icyph}
\eeq
\beq{}
i.c.B)~~x[0]=0,~r[0]=r_\mathrm{ini},~\dot{r}[0]=-\Big[{C'_r}^{_{\scriptstyle2}}h\Big]^\frac{1}{2}_{r=_\mathrm{ini}}
\label{icxph}
\eeq

\begin{figure*}[ht]
\begin{center}
\parbox{0.43\textwidth}{
\includegraphics[scale=0.6]{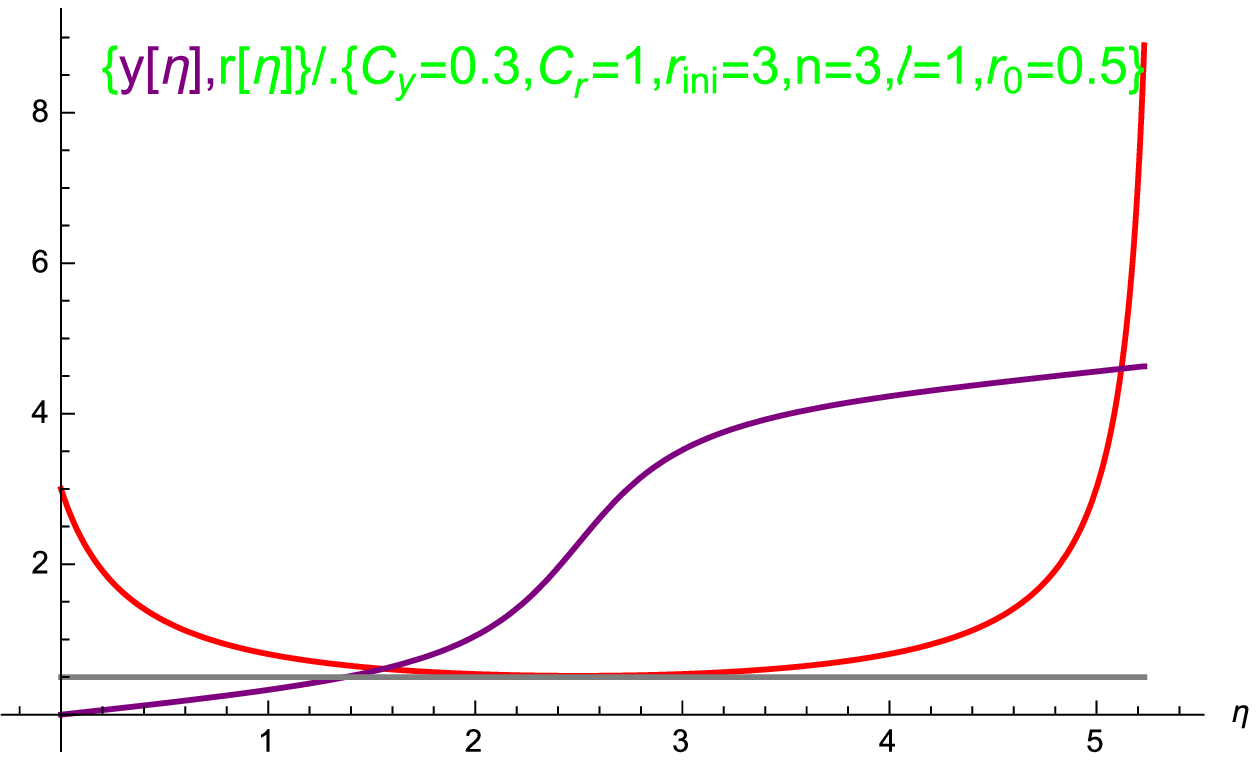}
\\\rule{3mm}{0pt}\includegraphics[scale=1]{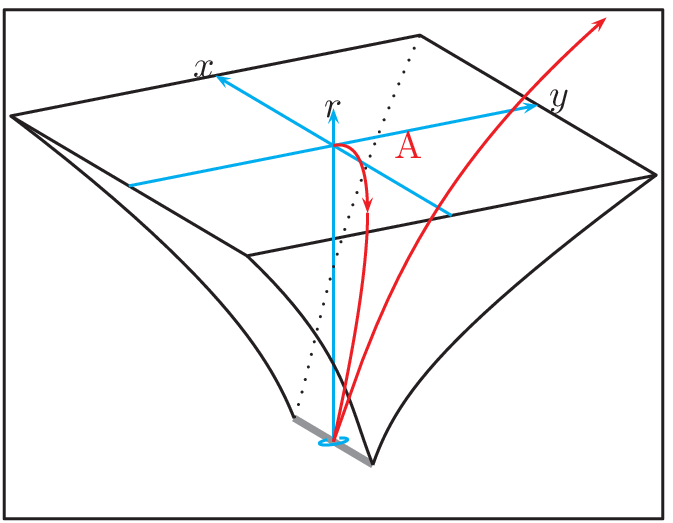}
}\parbox{0.43\textwidth}{
\includegraphics[scale=0.6]{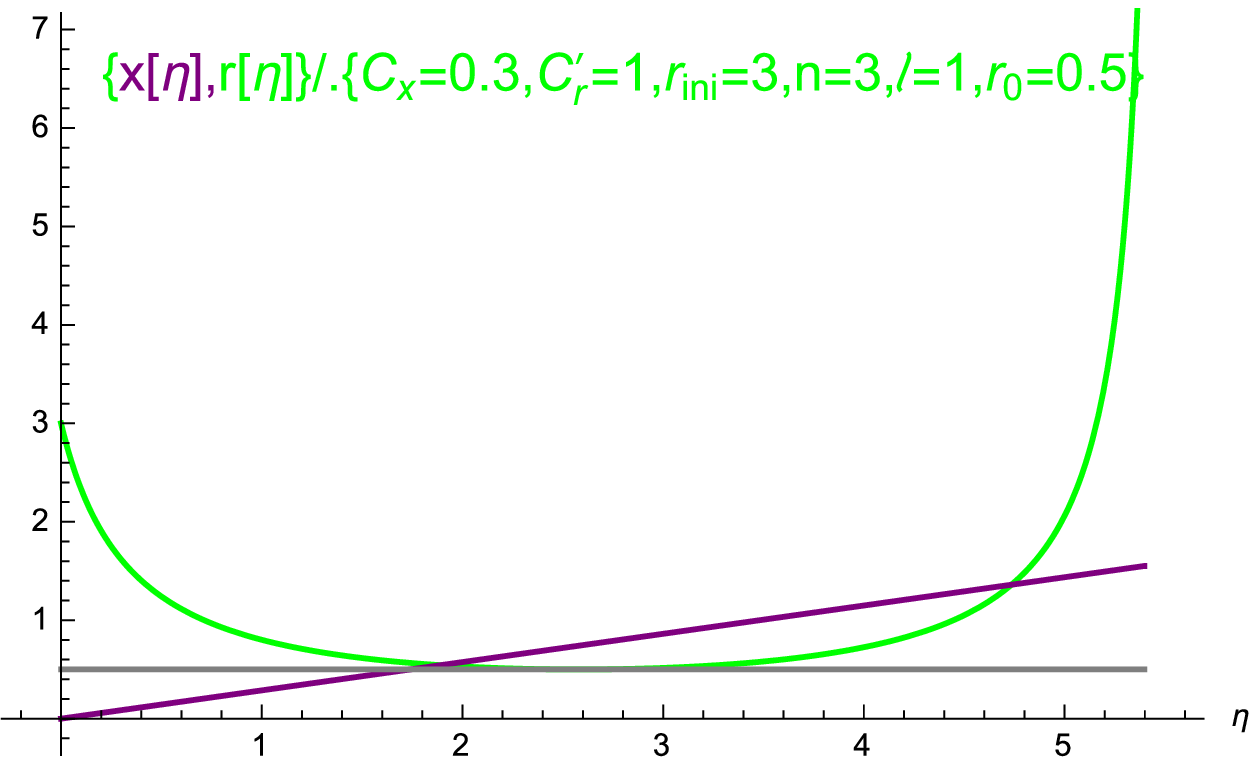}
\\\rule{3mm}{0pt}\includegraphics[scale=1]{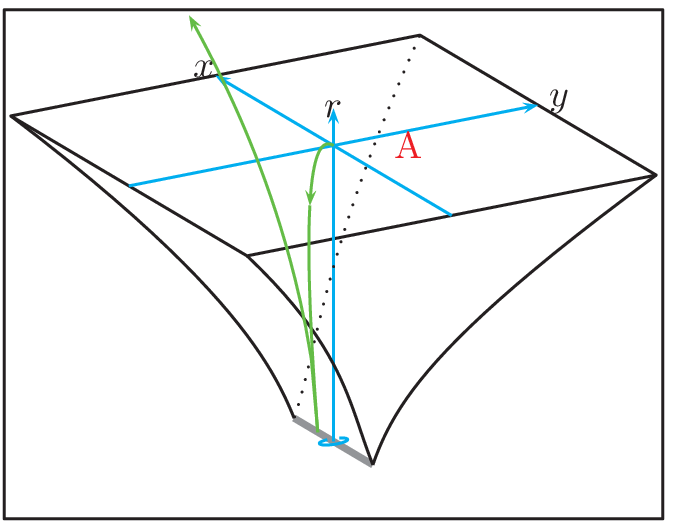}
}\caption{left, numeric solutions to the geodesic equation \eqref{eqYrph1} of massless particles in AdS-Sol background eq\eqref{solMetric} with the initial condition \eqref{icyph}. The particles move in the $y-r$ plane and bounces as they approaching the ``horizon'' of the AdS-Sol geometry. Right, numeric solutions to \eqref{eqXrph1}+\eqref{icxph}, the relevant particles moving in the $x-r$ plane but also bounces as they approaching the ``horizon'' line. All parameters involved has been displayed in the figure.}
\label{figMPhfunction}
\end{center}
\end{figure*}

Our numerical results are displayed in figure \ref{figMPhfunction}. From the figure, we easily see that, in both cases, the massless particles bounces as they move close to the ``horizon'' line of the AdS-Sol space-time. This again tells us that, the negative mass of AdS-Sol's produce anti-gravity and repel even massless particles approaching to them. What's more, we numerically observe that, as the mass of the AdS-Sol increases, the ingoing massless particles will be more rapidly repelled from them.

\section{Applications}
\label{secCosmo}
\begin{figure*}[ht]
\begin{center}
\includegraphics[scale=0.6]{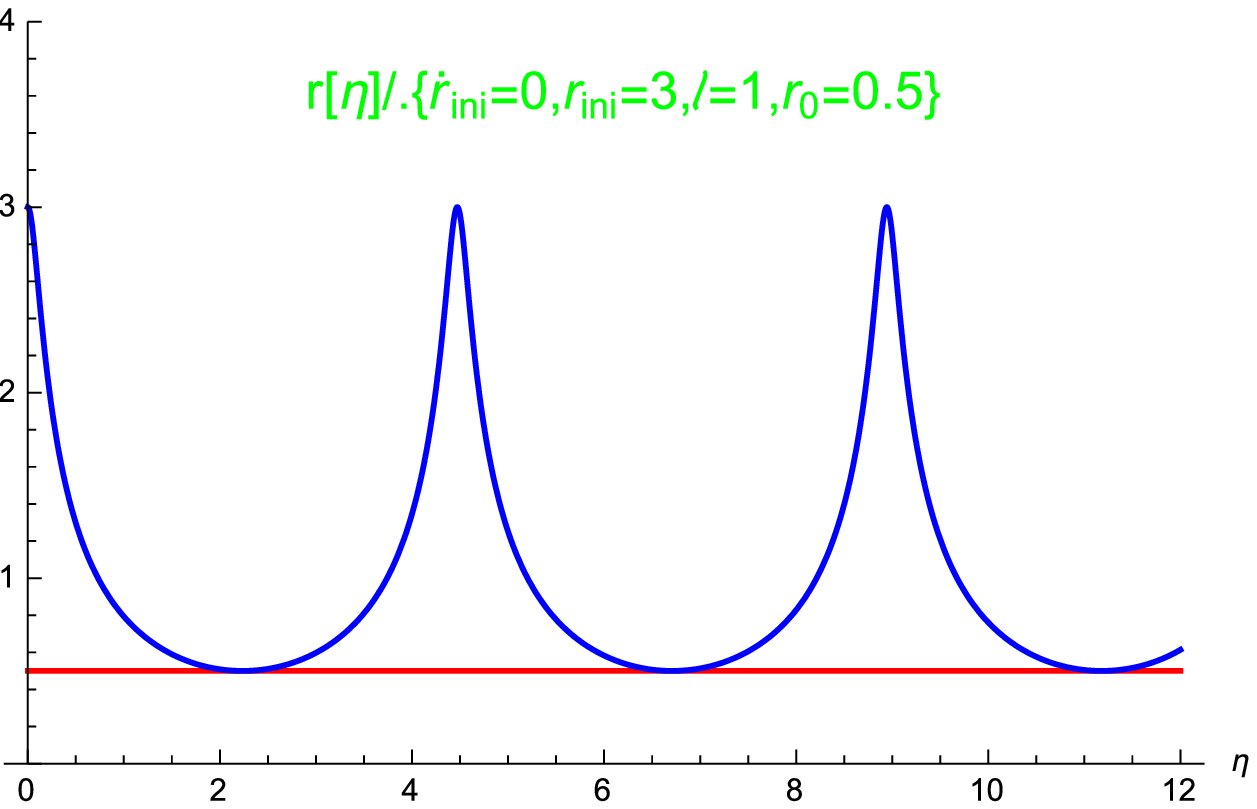}
\rule{2mm}{0pt}\includegraphics[scale=1]{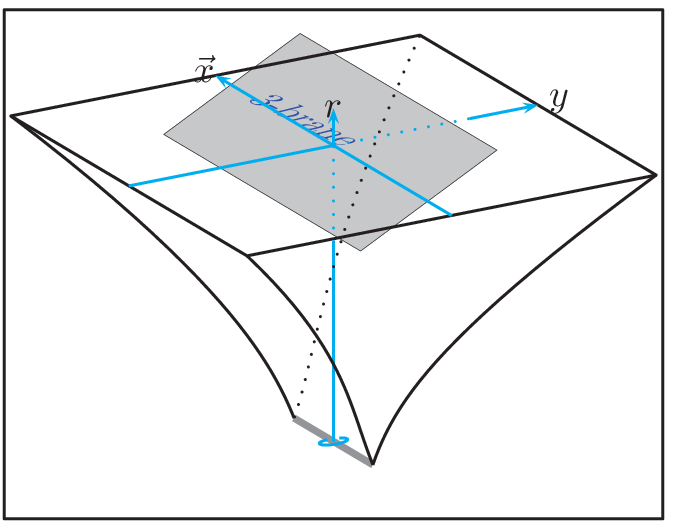}
\caption{A typical numerical solution to the radial motion of a 3+1-brane in (5+1)D AdS-Sol background. The brane occupies the $\eta,\vec{x}$-dimension of the background space-time. The corresponding brane-world cosmology bounces at the ``horizon'' of the soliton. Change the parameters $r_\mathrm{ini}$, $r_0$, $\dot{r}_\mathrm{ini}$/$C$ does not change this oscillating feature.}
\label{figRCosmic}
\end{center}
\end{figure*}

Facing with the anti-gravitation of AdS-Sol, one's first reaction may be, has it anything with the dark energy in the universe? Proper answer to this question needs us to study the anti-gravity (if appears) of general gravitation-solitons in asymptotically deSitter or Minkowski space-time. However, even limiting in AdS-Sol's, the facts observed in this paper provide us a very natural way of bouncing cosmological models building. Consider of a 3-brane occupying $\{\eta,\vec{x}\}$-dimensions and moving along the $r$-direction of a 6-dimensional AdS-Sol bulk space-time,
\bea{}
\textrm{AdS-Sol}_6:&&\hspace{-5mm}~ds^2=\frac{r^2}{\ell^2}(+hdy^2+d\vec{x}\cdot d\vec{x}-d\eta^2)+\frac{\ell^2dr^2}{r^2h}
\label{sol6Metric}
\\
\textrm{3-brane}:&&\hspace{-5mm}~\{\eta=\xi^0,\vec{x}=\vec{\xi},y\equiv0,r=r(\eta)\}
\eea
Obviously, the induced metric on the 3-brane has the standard of form Friedmann-Lematire-Robertson-Walker metrics in conformal time
\beq{}
ds_{3+1}^2=\frac{r(\eta)^2}{\ell^2}(-d\eta^2+d\vec{x}\cdot d\vec{x})
\eeq
with the scale factors now replaced by the radio coordinate $\frac{r(\eta)^2}{\ell^2}$. Its function determines evolutions of the brane-world cosmology. In the probe limit, backreactions of the brane on the background could be neglected and the dynamic follows from the simple Nambu-Goto action
\bea{}
S_\textrm{3-brane}&&\hspace{-5mm}=
\int\!d\eta\,d\vec{x}\sqrt{\det|\partial_aX^\mu\partial_bX^\nu G_{\mu\nu}|}
\\
&&\hspace{-5mm}=\int\!d\eta\,d\vec{x}\sqrt{\Big(1-\frac{\ell^4\dot{r}^2}{r^4h}\Big)\frac{r^8}{\ell^8}}
\eea
with the corresponding equations of motion reads
\beq{}
\frac{r^4/\ell^4}{\sqrt{1-\frac{\ell^4\dot{r}^2}{r^4h}}}=C
\eeq
where $C$ is a simple integral constant. Similar to the case in massive particles in section \ref{secMassive}, solutions to this equation has oscillating features. We plotted a typical solution in figure \ref{figRCosmic}.  From the figure we easily see that, the brane-world cosmology bounces as it evolves to the minimal size $\frac{r(\eta)}{\ell}=\frac{r_0}{\ell}$, with $r_0$ being the size of the bulk AdS-Sol geometry.

\section{Conclusion}
\label{secConclusion}
In this paper we firstly review the standard definition and calculation of energy/masses leading to the AdS-Sol geometry, which is negative but direct understanding is missing until this work. Then by studying geodesic motion of both massive and massless particle in the background of general dimensional AdS-Sol space-time, we provide an intuitive understanding of this negative energy/masses in pure gravitation theories instead of dual gauge theories. That is, they anti-gravitate. Under the potential-well effects of AdS-space-time and the repelling  the soliton's anti-gravitation, the massive particles oscillate along the radial direction, while massless particles experience one-time bouncing as they approach the ``horizon'' line of the soliton. Finally, as a potential application, we study the moving of a 3+1-brane in the background of 5+1 dimensional AdS-Sol background space-time in probe limit, we find that the brane also oscillates in the radial direction of the AdS-Sol. This forms a very natural model of bouncing/cyclic universe model.

\providecommand{\href}[2]{#2}\begingroup\raggedright

\end{document}